\documentclass[a4paper]{jpconf}
\usepackage{graphicx}
\begin{document}
\title{Wave function of the Universe, preferred reference frame effects and metric signature
transition}

\author{HOSSEIN GHAFFARNEJAD}

\address{Faculty of Physics, Semnan University, Semnan, IRAN, Zip Code: 35131-19111}

\ead{hghafarnejad@yahoo.com}

\begin{abstract}
Gravitational model of non-minimally coupled Brans Dicke (BD) scalar
field $\phi$ with dynamical unit time-like four vector field is used
to study flat Robertson Walker (RW) cosmology in the presence of
variable cosmological parameter $V(\phi)=\Lambda\phi.$  Aim of the
paper is to seek cosmological models which exhibit metric signature
transition. The problem is studied in both classical and quantum
cosmological approach with large values of BD parameter $\omega>>1$.
Scale factor of RW metric is obtained as
$R(t)=6\sqrt{\frac{3}{\Lambda}}\cosh\big(\frac{t}{4}\sqrt{\frac{\Lambda}{3}}\big)$
which describes nonsingular inflationary universe in Lorentzian
signature sector. Euclidean signature sector of our solution
describes a re-collapsing universe and is obtained from analytic
continuation of the Lorentzian sector by exchanging $t\to it$ as
$R(t)=6\sqrt{\frac{3}{\Lambda}}\cos\big(\frac{t}{4}\sqrt{\frac{\Lambda}{3}}\big).$
 Dynamical vector field together with the BD scalar field are treated as fluid
  with time dependent barotropic index. They have regular (dark)
matter dominance in the Euclidean (Lorentzian) sector. We solved
Wheeler De
  Witt (WD) quantum wave equation of the cosmological system. Assuming a discrete non-zero ADM
mass $M_j=4\sqrt{2}(2j+1)\sqrt{\frac{\Lambda}{3}}$ with
$j=0,1,2,\cdots$, we obtained solutions of the WD equation as simple
harmonic quantum Oscillator eigen functionals described by Hermite
polynomials. Absolute values of these eigen functionals have nonzero
values on the hypersurface $R=6\sqrt{\frac{3}{\Lambda}}$ in which
metric field has signature degeneracy. Our eigen functionals
describe nonzero probability of the space time with Lorentzian
(Euclidean) signature for $R>6\sqrt{\frac{3}{\Lambda}}$
($R<6\sqrt{\frac{3}{\Lambda}}$). Maximal probability corresponds to
the ground state $j=0.$
\end{abstract}
\section{Introduction}
Lorentz invariance is a fundamental requirement of the standard
model of particle physics, verified to high precision by many tests
[1], whereas, string theory predicts that we may live in a universe
with non-commutative coordinates [2] which leads to a violation of
Lorentz invariance [3]. Furthermore, astrophysical observations
suggest the presence of high energy cosmic rays above the
Greisen-Zatsepin-Kuzmin cutoff [4] which may be explained by a
breaking down of Lorentz invariance. More generally, Lorentz
invariance violation seems to be related to unknown physical effects
of space time at the Planck scale $(l_p=(16\pi G)^{-1/2}$ with
$c=\hbar=1$). A straightforward method of implementing local Lorentz
violation in a gravitational setting is to introduce a tensor field
with a non-vanishing expectation value, and then couple this tensor
to gravity or matter fields. The simplest proposal of this approach
is to consider a single time-like vector field $N_{\mu}$ with a
fixed norm [5]. Physical interpretation of such a vector field
denotes to gravitational effects of preferred reference frames.
Regarding general covariance principle, $N_{\mu}$ should be taken as
a dynamical field. There have been also elsewhere, cosmological
studies in the presence of vector fields without the fixed norm.
Lorentz invariance violation  directly causes the change of the
metric signature of space time. The motivation for the metric
signature transition is usually caused by the extension of real
functions to corresponding functions with complex variables via
trivial Wick rotations in a complex plane.
 According to the work presented by Barbero et al [5], the author's attempt was previously
 to generalize the BD gravity
action [6,7] by transforming its background metric as $g_{\mu\nu}\to
g_{\mu\nu}+2N_{\mu}N_{\nu}$
 where $N_{\mu}$ is assumed to be a dynamical unit time like vector
 field.
Further, dust,
 radiation and inflationary
cosmological models were obtained.\\
 In this lecture we first present our scalar-vector-tensor gravity [6,7]
 in the presence of variable cosmological
 parameter. Then we seek flat RW cosmological models which exhibit metric
 signature transition from Lorentzian (-,+,+,+) to Euclidean (+,+,+,+) sectors
 in both classical and quantum approaches. Last section contains the concluding remark.
\section{Brans-Dicke scalar-vector-tensor Robertson Walker cosmology}
Let us take the scalar-vector-tensor-gravity action  $
I_{total}=I_{BD}+I_{\Lambda}+I_{N}$ [6,7] where
  the first term is well known BD scalar tensor gravity [8] as
 $ I_{BD}=\frac{1
 }{16\pi}\int dx^4\sqrt{g}\left\{\phi R-\frac{\omega}{
 \phi}
 g^{\mu\nu}\nabla_{\mu}
 \phi\nabla_{\nu}\phi\right\},$ the second term
 denotes the action of self interacting BD field (called as
`variable cosmological parameter`) such that $
I_{\Lambda}=\frac{\Lambda}{16\pi}\int dx^4\sqrt{g}\phi^n,$
 in which $\Lambda,n$ are real constant parameters
and the last term given by $I_N=\frac{1}{16\pi}\int
 dx^4\sqrt{g}\{\zeta(x^{\nu})(g^{\mu\nu}N_{\mu}N_{\nu}+1)+2\phi F_{\mu\nu}F^{\mu\nu}-\phi N_\mu
 N^{\nu}(2F^{\mu\lambda}\Omega_{\nu\lambda}+
 F^{\mu\lambda}F_{\nu\lambda}+\Omega^{\mu\lambda}\Omega_{\nu\lambda}
-
2R_{\mu}^{\nu}+\frac{2\omega}{\phi^2}\nabla_{\mu}\phi\nabla^{\nu}\phi)\}$
with $F_{\mu\nu}=2(\nabla_{\mu}N_{\nu}-\nabla_{\nu}N_{\mu}),$ and
$\Omega_{\mu\nu}=2(\nabla_{\mu}N_{\nu}+\nabla_{\nu}N_{\mu})$,
describes action of unit time like dynamical four vector field
$N_{\mu}(x^{\nu}).$ Non-minimally vector field $N_{\mu}$ is called
usually as four velocity of preferred reference frame. The action
$I_{total}$ is written in units $c=\hbar=1$ with Lorentzian
signature (-,+,+,+). The undetermined Lagrange multiplier
$\zeta(x^{\nu})$ controls that $N_{\mu}$ to be an unit time-like
vector field. Self-interacting coupling constant $\Lambda$ has
dimension as $(length)^{2n-4}.$  Absolute value of determinant of
the metric $g_{\mu\nu}$ is defined by $g$. BD field $\phi$ describes
inverse of variable Newton`s gravitational coupling parameter and
its dimension is $(length)^{-2}$ in units $c=\hbar=1$. Present
limits of $\omega$ based on time-delay experiments [9,10,11] require
$\omega\geq4\times10^{4}$ and general relativistic approach $I_{BD}$
is obtained with $\omega\to\infty$ [12,13,14]. As an example we set
RW line element which with Lorentzian signature (-,+,+,+) is given
from point of view of free falling comoving observer as
$ds^2=-dt^2+R^2(t)\{dx^2+dy^2+dz^2\}$ where $R(t)$ is spatially part
scale factor of space time. In this case one can obtain explicitly
time dependent fields $\zeta(t),$ $\phi(t)$ and $R(t)$ such as
follows: $\phi(t)=\phi_0 \exp[8/(2\omega-5)(\Omega t)]$ and
$R(t)=R_0\cosh(\Omega t)$ (details of the mathematical calculations
are given in ref. [15]), where
 $\phi_0=\phi(0),R_0=R(0)$ and we defined
$\Omega=\frac{3(2\omega-5)}{4\sqrt{(6\omega-3-2\sqrt{6})(6\omega-3+2\sqrt{6})}}\sqrt{\frac{\Lambda}{3}}.
$ The parameter $\Omega$ has real value under the conditions
$\frac{1}{2}-\frac{\sqrt{6}}{3}<
\omega<\frac{1}{2}+\frac{\sqrt{6}}{3}$ with $\Lambda<0$ or
$\frac{1}{2}-\frac{\sqrt{6}}{3}>
\omega>\frac{1}{2}+\frac{\sqrt{6}}{3}$ with $\Lambda>0.$ Our
obtained scale factor describes non-singular inflationary expanding
flat universe. With the above obtained solutions the undetermined
Lagrange multiplier $\zeta(t),$ fluid density $\rho(t)$ and
corresponding pressure $p(t)$ become respectively
$\zeta(t)=\bigg\{\frac{(6\omega-31-4\sqrt{46})(6\omega-31+4\sqrt{46})}
{(2\omega-5)(6\omega-3-2\sqrt{6})(6\omega-3+2\sqrt{6})}\bigg\}\Lambda\phi_0\exp(\Omega
t)$, $\rho(t)=3\Omega^2\tanh^2\Omega t,$ and
$p(t)=-\Omega^2\{2+\tanh^2\Omega t\}$ where we set initial condition
as $\rho(0)=0.$
  One can obtain equation of
state as $
\eta(t)=\frac{p(t)}{\rho(t)}=-\frac{1}{3}-\frac{2}{3}\coth^2\Omega
t$ which describes dark matter dominant [16] of the fluid from point
of view of free falling comoving frame. It is applicable to choose a
preferred reference frame with its corresponding time coordinate $T$
related to the cosmological comoving time $t$ as $ T(t)=(\Omega
t)^{\frac{2}{3}}.$ From the point of view of the latter preferred
reference frame the RW line element can be rewritten as
$ds^2=R_0^2d\bar{s}^2$ in which
$d\bar{s}^2=-TdT^2+\cosh^2(T^{3/2})\{dx^2+dy^2+dz^2\}$ and we set
$R_0=\frac{3}{2\Omega}.$ We see that hypersurfaces $t=constant$ with
$T>0$ correspond to Lorentzian signature (-,+,+,+) of the space time
metric. Its analytic continuation is obtained with hypersurfaces
$it=constant$ which correspond to Euclidean signature (+,+,+,+) with
negative times $T<0.$ Hence metric components have degeneracy on the
signature transition hypersurface $T=0$ (see LHS in the figure 1).
Diagram of the barotropic index $\eta(T)$ is plotted against $T$ in
 LHS of the figure 1 with dash line.
\begin{figure}
\begin{center}\includegraphics[width=16pc]{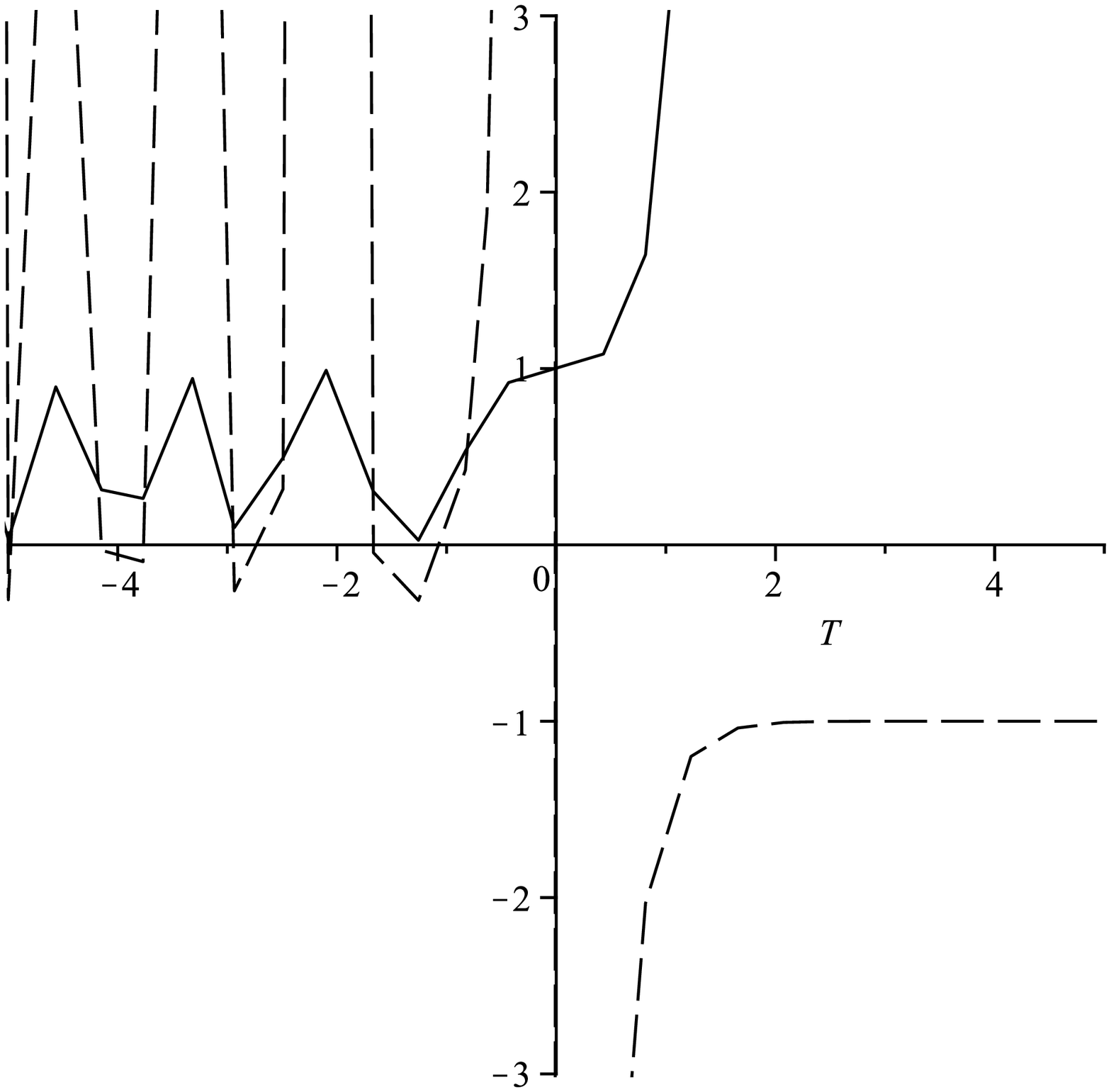}
\includegraphics[width=16pc]{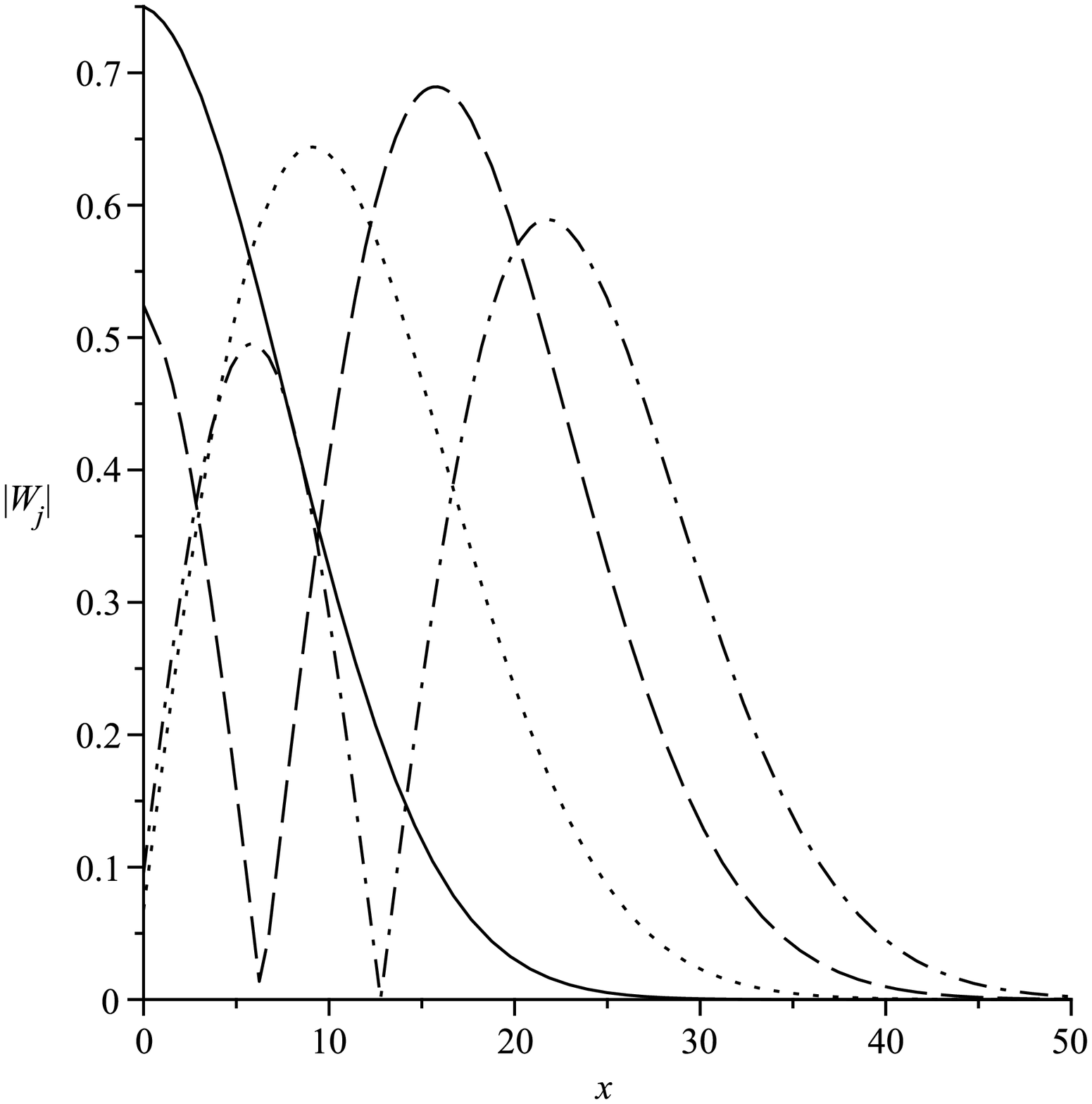}
\caption{\label{label} Solid line in left diagram describes
Lorentzian (Euclidean) sector scale factor of RW space time for
times $T>0(T<0).$ Dash line in left diagram describes barotropic
index $\eta(T)$ of Lorentzian (Euclidean) sector for times
$T>0(T<0)$. Right diagram describes variation of absolute value of
the WD eigen functionals $|W_j(x)|$ against scale factor $x$ for
$j=0,1,2,3$ with solid, dot, dash and dash-dot lines respectively.}
\end{center}\end{figure} It shows that the scalar-vector matter
source treats as regular (dark) matter in the Euclidean (Lorentzian)
sector of the space time. In the quantum level we obtain below that
the Lorenzian sector of space times  may be tunneled quantum
mechanically to the corresponding Euclidean sector and vice versa.
Using minisuperspace approach of canonical quantum cosmology one can
obtain corresponding Wheeler DeWitt probability wave equation of the
system as $ \{\frac{(6\omega^2-81\omega-27)}{4(2\omega+3)^2}
 \frac{\delta^2}{\delta R^2}+\frac{3(7\omega+12)}{(2\omega+3)^2}\frac{\phi}{R}
 \frac{\delta}{\delta R}\frac{\delta}{\delta\phi}-\frac{(2\omega+21)}{2(2\omega+3)^2}
 \frac{\phi^2}{R^2}\frac{\delta^2}{\delta\phi^2}
 -\phi R(M+\Lambda\phi R^3)\}W(R,\phi)=0$
(details of mathematical calculations are given in ref. [15]). In
general relativistic approach $\omega\to+\infty$ our classical
solution leads to $\lim_{\omega\to+\infty}\phi(t)=\phi_0=constant$
and
$\lim_{\omega\to+\infty}R(t)=R_0\cosh\bigg(\frac{t}{4}\sqrt{\frac{\Lambda}{3}}
 \bigg)$ where $R_0=\frac{3}{2\Omega}=6\sqrt{\frac{3}{\Lambda}}$ and $\phi_0=\frac{1}{16\pi
 G}.$ The above WD wave equation has not a simple solution
 with small $\omega,$ but we can  obtain some physical interpretation from solution of
 its general
 relativistic approach $\omega\to+\infty.$ In the latter approach the WD
 wave equation becomes $W''_{\infty}+U(x)W_{\infty}(x)\cong0$ where we defined
$U(x)=-x(a+bx^3),$ $a=\frac{108M}{\pi
G\Lambda}\sqrt{\frac{3}{\Lambda}}$, $b=\frac{1944}{\pi^2
G^2\Lambda^2}$, $x=\frac{R}{R_0}$ and over-prime $^{\prime}$ denotes
differentiation with respect to the minisuperspace variable $x$.
Subscript of the WD wave solution $W_{\infty}$ denotes to
$\omega\to+\infty.$ We point that $R_0$ is scale factor of space
time where the metric has signature degeneracy.
  Solving $U^{\prime}(x)=0$ we obtain
minimum point of the potential $U(x)$ as
$x_m=-\frac{1}{2}\bigg(\frac{\pi GM}{3}\sqrt{\frac{\Lambda}{3}}
\bigg)^{1/3}$ for which corresponding Taylor series expansion
becomes $U(x)\cong \frac{81}{2}\bigg(\frac{M}{\Lambda\sqrt{\pi
G}}\bigg)^{\frac{4}{3}}-\frac{1944}{\Lambda}\bigg(\frac{M}{\pi^2G^2\Lambda}
\bigg)^{\frac{2}{3}}(x-x_m)^2+\cdots .$ Defining $x-x_m=\sigma y$
where
$\sigma=\bigg(\frac{\Lambda}{1944}\bigg)^{\frac{1}{4}}\bigg(\frac{\pi^2
G^2\Lambda}{M}\bigg)^{\frac{1}{6}}$ the WD wave equation becomes
$W^{\prime\prime}(y)+(\varepsilon-y^2)W(y)=0$ where we eliminated
subscript $\infty$ and defined
$\varepsilon=\frac{M}{4\sqrt{2}}\sqrt{\frac{3}{\Lambda}}.$ This
equation describes simple harmonic quantum Oscillator and its
solutions are described in terms of Hermite polynomials
$H_j(y)=(-1)^j e^{y^2}\frac{d^j}{dy^j}e^{-{y^2}}$ as $
W_j(y)=\left(\frac{1}{2^jj!\sqrt{\pi}}\right)^{\frac{1}{2}}\exp(-\frac{y^2}{2})H_j(y)$
with eigen values $\varepsilon_j=(2j+1),~j=0,1,2,\cdots .$ In terms
$x$ our obtained WD eigen functions become
$W_j(x)=\left(\frac{1}{2^jj!\sqrt{\pi}}\right)^{\frac{1}{2}}\exp\{-\frac{\sigma_j^2(x-x_m^j)^2}
{2}\}H_j(\sigma_j(x-x_m^j))$ in which
$x_m^j=-\left(\frac{\pi\sqrt{2}(2j+1)\Lambda
G}{18}\right)^{\frac{1}{3}}$,
$\sigma_j=\bigg(\frac{1}{1944}\bigg)^{\frac{1}{4}}\bigg(\frac{\pi^2\Lambda^2
G^2}{4(2j+1)}\sqrt{\frac{3}{2 }}\bigg)^{\frac{1}{6}}$ and discrete
ADM mass becomes $ M_j=4\sqrt{2}(2j+1)\sqrt{\frac{\Lambda}{3}}.$
Their superposition leads to general wave solution of quantum
universe of our model as $W(x)=\sum_{j=0}^{\infty}P_jW_j(x)$ where
$P_j$ is probability amplitude of the universe which should stay on
the eigen state $j$ with ADM eigen mass (energy) $M_j$. However our
knowledge about the universe is not enough and so we can not
determine the coefficients $P_j$, because LHS in the above summation
$`W(x)`$ is unknown. But we can obtain some physical statements
about the quantum tunneling of metric signature
 via eigen states $W_j(x)$: Our classical solution
$x(T)=\frac{R}{R_0}=\cosh(T^{3/2})$ predicts a metric signature
transition on the hypersurface $T=0$ corresponding to $x=1.$
Lorentzian sector of space time corresponds to $x>1$ but its
Euclidean sector corresponds to $0<x<1$ (see LHS  of the figure 1).
The metric equation has signature degeneracy on the hypersurface
$x=1$ corresponding to the time $T=0$ in the classical regime.
$W_j(0<x<1)$ $(W_j(x>1))$ describe the probability of Euclidean
(Lorentzian) signature of the space time. Nonzero values
$W_j(x=1)\neq0$ describe the probability of metric signature quantum
tunneling which takes maximal value when the universe exhibits its
ground state $j=0$ (see RHS of the figure 1 where we set $\Lambda
G\pi=1$).
\section{Summary and concluding remarks} Flat RW space time is studied by using
the BD scalar-vector tensor gravity and variable cosmological
parameter. A nonsingular inflationary cosmological model is obtained
which exhibits metric signature transition from Lorentzian to
Euclidean topology. Unit time like dynamical vector field and BD
scalar field are treated as regular (dark) fluid in Euclidean
(Lorentzian) sector of flat RW space time with time dependent
barotropic index. WD wave equation of the cosmological system is
solved by assuming nonzero ADM mass. Its solutions are obtained as
simple harmonic quantum Oscillator eigen functionals. Maximal
probability of signature quantum tunneling is obtained when the
quantum Oscillating RW universe lives in its ground state. As future
work the author intends to investigate the dynamical effects of
preferred reference frames on black holes and
anisotropic-inhomogeneous Bianchi cosmological metrics .

\smallskip

\end{document}